\documentclass[3p,times,procedia]{elsarticle}
\usepackage{nupha_ecrc}

\volume{00}

\firstpage{1}

\journalname{Nuclear Physics A}

\runauth{Javier L. Albacete et al.}

\jid{nupha}

\jnltitlelogo{Nuclear Physics A}

\usepackage{amssymb}
\usepackage[figuresright]{rotating}

\begin{document}

\begin{frontmatter}



%% Instructions from Editor: Please use the following \dochead only in the preprint version (e-print arXiv etc.); 
%% use empty \dochead{} when submitting to Nuclear Physics A!
\dochead{XXVIIth International Conference on Ultrarelativistic Nucleus-Nucleus Collisions\\ (Quark Matter 2018)}
%% e.g. \dochead{17th International Conference on Dynamical Processes in Excited States of Solids}

\title{Correlated gluonic hot spots meet symmetric cumulants data at LHC energies}

\author[label1]{Javier L. Albacete}
\ead{albacete@ugr.es}
\address[label1]{CAFPE and Dpto. de F\'isica Te\'orica y del Cosmos, Universidad de Granada, E-18071 Campus de Fuentenueva, Granada, Spain.} 

\author[label2,label3,label4]{Harri Niemi}
\ead{harri.m.niemi@jyu.fi}

\address[label2]{Institut f{\"u}r Theoretische Physik, Johann Wolfgang Goethe-Universit{\"a}t, Max-von-Laue-Str. 1, D-60438 Frankfurt am Main, Germany.}

\address[label3]{Department of Physics, University of Jyv{\"a}skyl{\"a}, P.O. Box 35, FI-40014 University of Jyv{\"a}skyl{\"a}, Finland.}

\address[label4]{Helsinki Institute of Physics, P.O. Box 64, FI-00014 University of Helsinki, Finland.}

\author[label2,label5,label6]{Hannah Petersen}

\ead{petersen@fias.uni-frankfurt.de}

\address[label5]{Frankfurt Institute for Advanced Studies, Ruth-Moufang-Strasse 1, 60438 Frankfurt am Main, Germany.}

\address[label6]{GSI Helmholtzzentrum f{\"u}r Schwerionenforschung, Planckstr. 1, 64291 Darmstadt, Germany.}

\author[label1,label5]{Alba Soto-Ontoso}

\ead{ontoso@fias.uni-frankfurt.de}

\begin{abstract}

We present a systematic study on the influence of spatial correlations between the proton constituents, in our case gluonic hot spots, their size and their number on the symmetric cumulant SC$(2,3)$, at the eccentricity level, within a Monte Carlo Glauber framework~\cite{Albacete:2017ajt}. When modeling the proton as composed by 3 gluonic hot spots, the most common assumption in the literature, we find that the inclusion of spatial correlations is indispensable to reproduce the negative sign of SC$(2,3)$ in the highest centrality bins as dictated by data. Further, the subtle interplay between the different scales of the problem is discussed. To conclude, the possibility of feeding a 2+1D viscous hydrodynamic simulation with our entropy profiles is exposed.

\end{abstract}


\end{frontmatter}

% \linenumbers

\section{Motivation}
The relevance of subnucleonic degrees of freedom and their fluctuations in the description of multiple experimental observations in small collision systems, such as flow harmonics~\cite{Mantysaari:2017cni}, diffractive phenomena~\cite{Mantysaari:2016ykx} or the hollowness effect~\cite{Albacete:2016pmp}, has been recently established. Although the geometric structure of the proton is a model-dependent quantity, stringent constrains can be extracted by means of Bayesian techniques~\cite{Moreland:2017kdx} or by identifying experimental observables with large discriminating power. 

A representative example is the first measurement of symmetric cumulants, SC$(n,m)$, performed by the CMS Collaboration in the three collision systems available at the LHC (p+p, p+Pb, Pb+Pb)~\cite{Sirunyan:2017uyl}. In particular, SC$(2,3)$, that provides direct access to initial state fluctuations, shows a sign change when moving towards large multiplicities (N$_{\rm{ch}}\!\sim\!100$) in p+p resembling the behavior of p(Pb)+Pb interactions. While there are several theoretical calculations that confront the p+Pb~\cite{Dusling:2017aot} and Pb+Pb~\cite{Giacalone:2016afq} measurements, the p+p data set lacks, up to today, of a succesful theoretical description. In this work, largely based on \cite{Albacete:2017ajt}, a plausible explanation for the sign change of SC$(2,3)$ when enlarging the multiplicity in p+p collisions at $\sqrt s\!=\!13$~TeV is exposed. 

\section{Setup}\label{model}
\begin{figure}
\begin{minipage}{.5\textwidth}
  \centering
\includegraphics[width=1\textwidth]{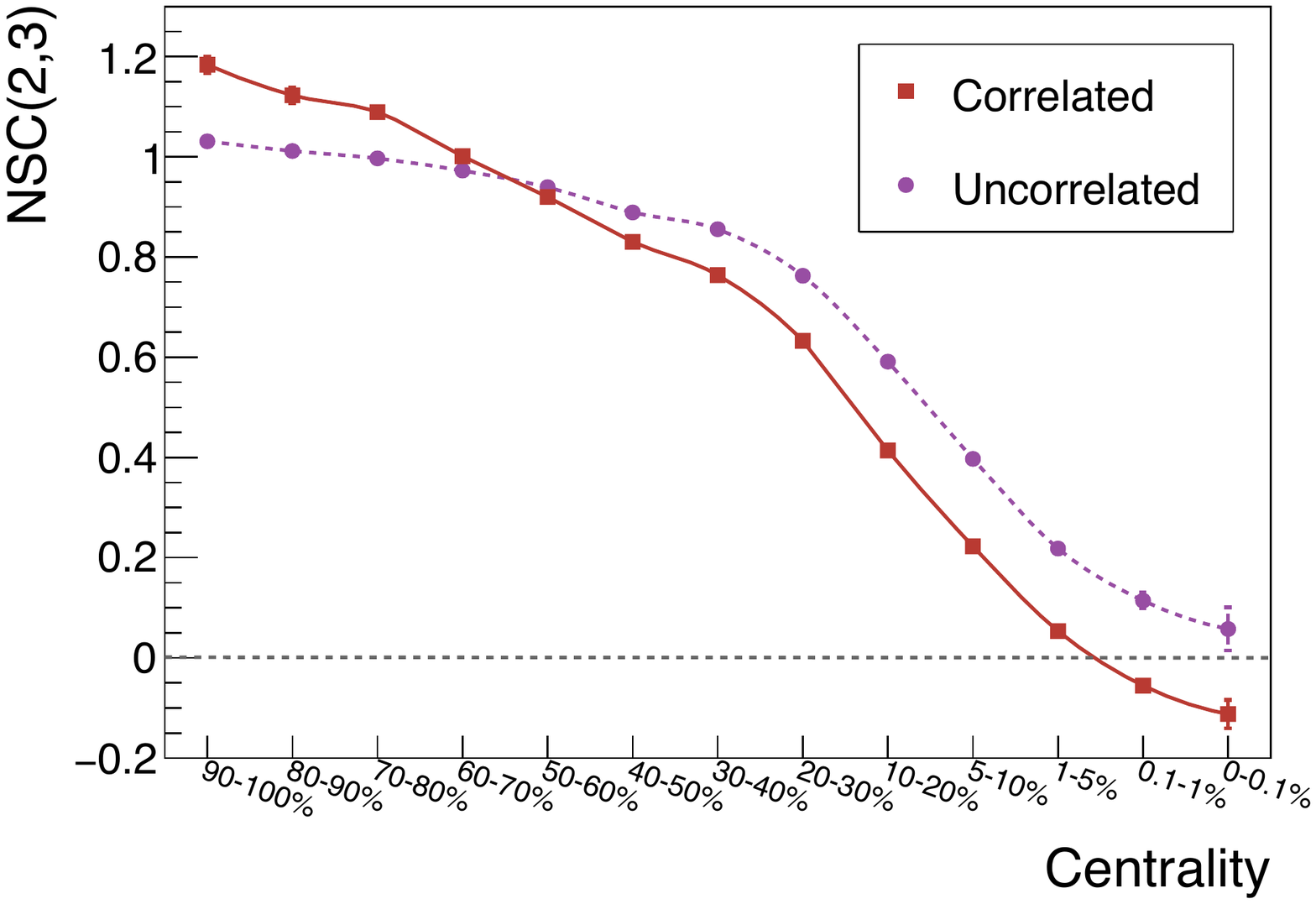}
\end{minipage}%
\begin{minipage}{.5\textwidth}
  \centering
\includegraphics[width=1\textwidth]{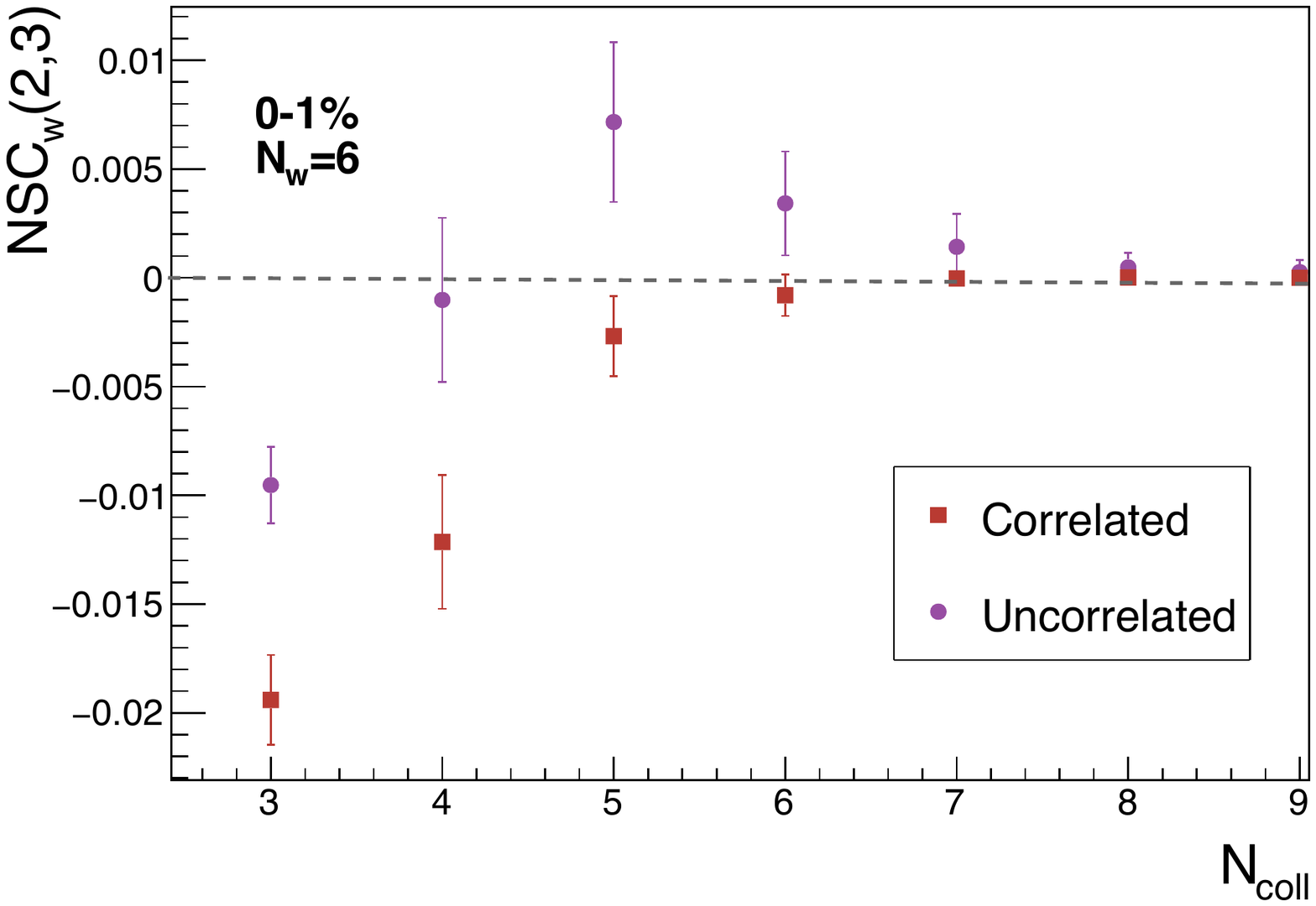}
\end{minipage}
\vspace*{-0.25cm}
\caption{Left: Average value of NSC$(2,3)$ as a function of the centrality range for the correlated (red solid line connecting filled red squares) and uncorrelated (purple solid line connecting filled purple circles). The error bars account for statistical uncertainties. Right: Average value of NSC$_w(2,3)$ as a function of the number of collisions after selecting the events with N$_w\!=\!6$ in the $[0\!-\!1\%]$ centrality bin for the uncorrelated (filled purple circles) and correlated (filled red squares) scenarios.}
\label{nsc23plot}
\end{figure}

A complete description of the model can be found in~\cite{Albacete:2016gxu}. Basically, we rely on a Monte Carlo Glauber description of the scattering process being the proton constituents, i.e. gluonic hot spots, the fundamental degrees of freedom. They are characterized by their radius $R_{hs}$ and their number $N_{hs}$, set by default to 3. An essential ingredient in any Glauber simulation is the spatial distribution of the constituents in the transverse plane. We introduce a novel ingredient with respect to other works in the literature (see Eq.2 in \cite{Albacete:2016gxu}): short-range repulsive correlations between the gluonic hot spots controlled by a repulsive core distance $r_c$ that effectively enlarge their mean transverse separation. Then, each wounded hot spot deposits, following a Gaussian distribution, a fluctuating amount of entropy tightly related with the event multiplicity. Moreover, we characterize the centrality of an event by its deposited entropy. With all these ingredients we compute the normalized symmetric cumulant, NSC$(2,3)$, at the eccentricity level defined as:
\begin{equation}
{\rm {NSC}}(2,3)=\displaystyle\frac{\langle\varepsilon_2^2\varepsilon_3^2\rangle-\langle\varepsilon_2^2\rangle\langle\varepsilon_3^2\rangle}{\langle\varepsilon_2^2\rangle\langle\varepsilon_3^2\rangle}
\end{equation}
in the correlated and uncorrelated scenarios. The values of the model parameters are given in Table 1 of \cite{Albacete:2017ajt}. 
\section{Results}
The main result of our analysis is presented on the left pannel of Fig.~\ref{nsc23plot} where we show the event-averaged value of NSC$(2,3)$ as a function of centrality. The most striking effect of the short-range repulsive correlations is observed in the ultra-central bins [$0\!-\!0.1\%$] and [$0.1\!-\!1\%$]: only in the correlated case there exists an anti-correlation of $\varepsilon_2$ and $\varepsilon_3$ as data dictates. Then, we conclude that the experimental evidence of NSC$(2,3)<\!0$ may back up the necessity to consider correlated proton constituents.

The reason why the correlations push NSC$(2,3)$ to negative values is displayed on the right pannel of Fig.~\ref{nsc23plot}. We compute NSC$(2,3)$ for a given number of wounded hot spots, $N_w$, weighted by the probability of these configurations to happen in the Monte Carlo, NSC$_w$(2,3), as a function of the number of collisions $N_{\rm {coll}}$. Clearly, the events with a large number of wounded hot spots and a small number of collisions are responsible for the negative sign of NSC$(2,3)$ within our approach. The main role of the correlations is just to enhance the probability of occurrence of these interaction topologies as compared to the uncorrelated scenario in the Monte Carlo and allow the change of sign. 

\begin{figure}
\begin{minipage}{.5\textwidth}
  \centering
\includegraphics[width=1\textwidth]{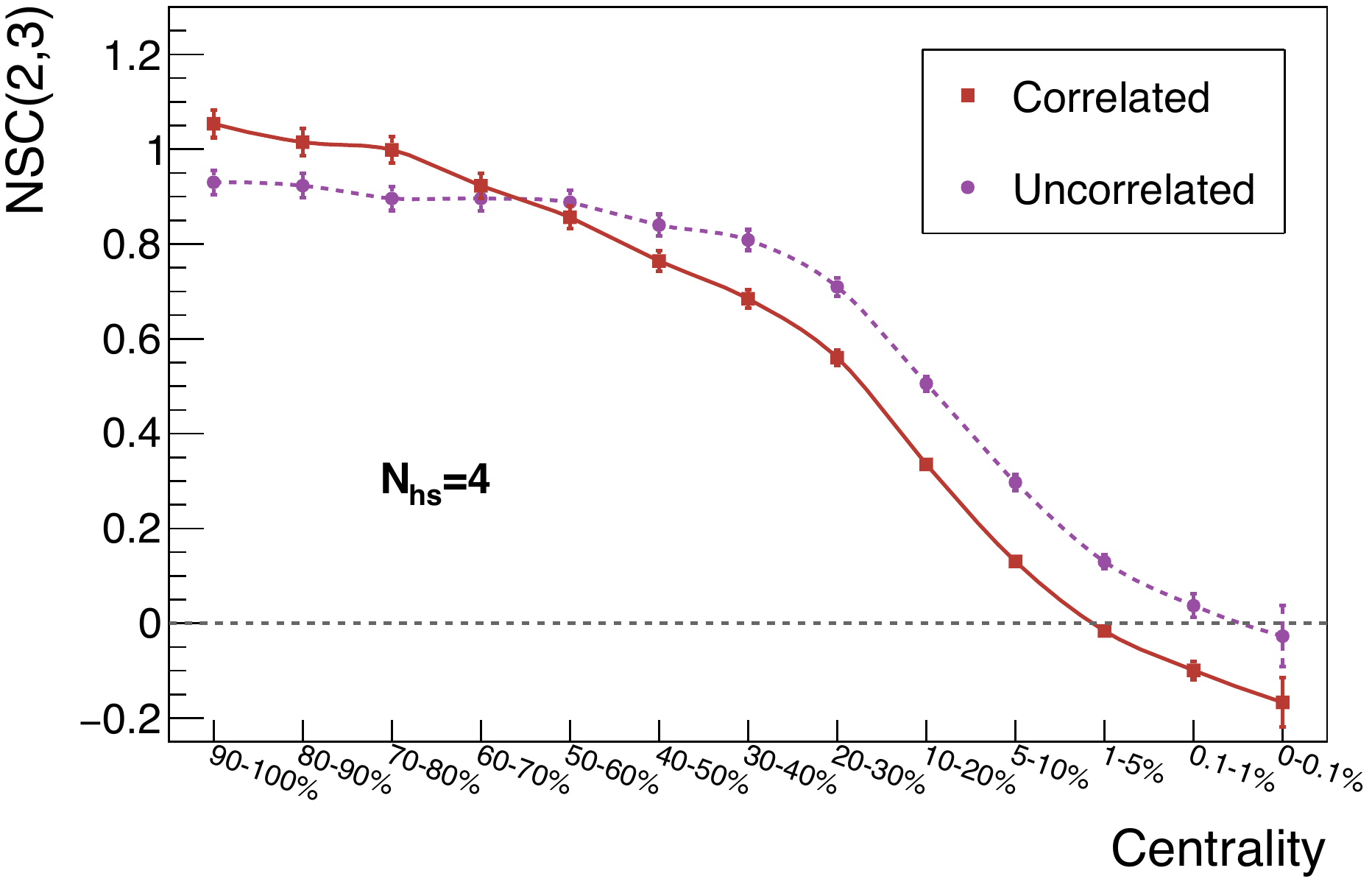}
\end{minipage}%
\begin{minipage}{.5\textwidth}
  \centering
\includegraphics[width=0.9\textwidth]{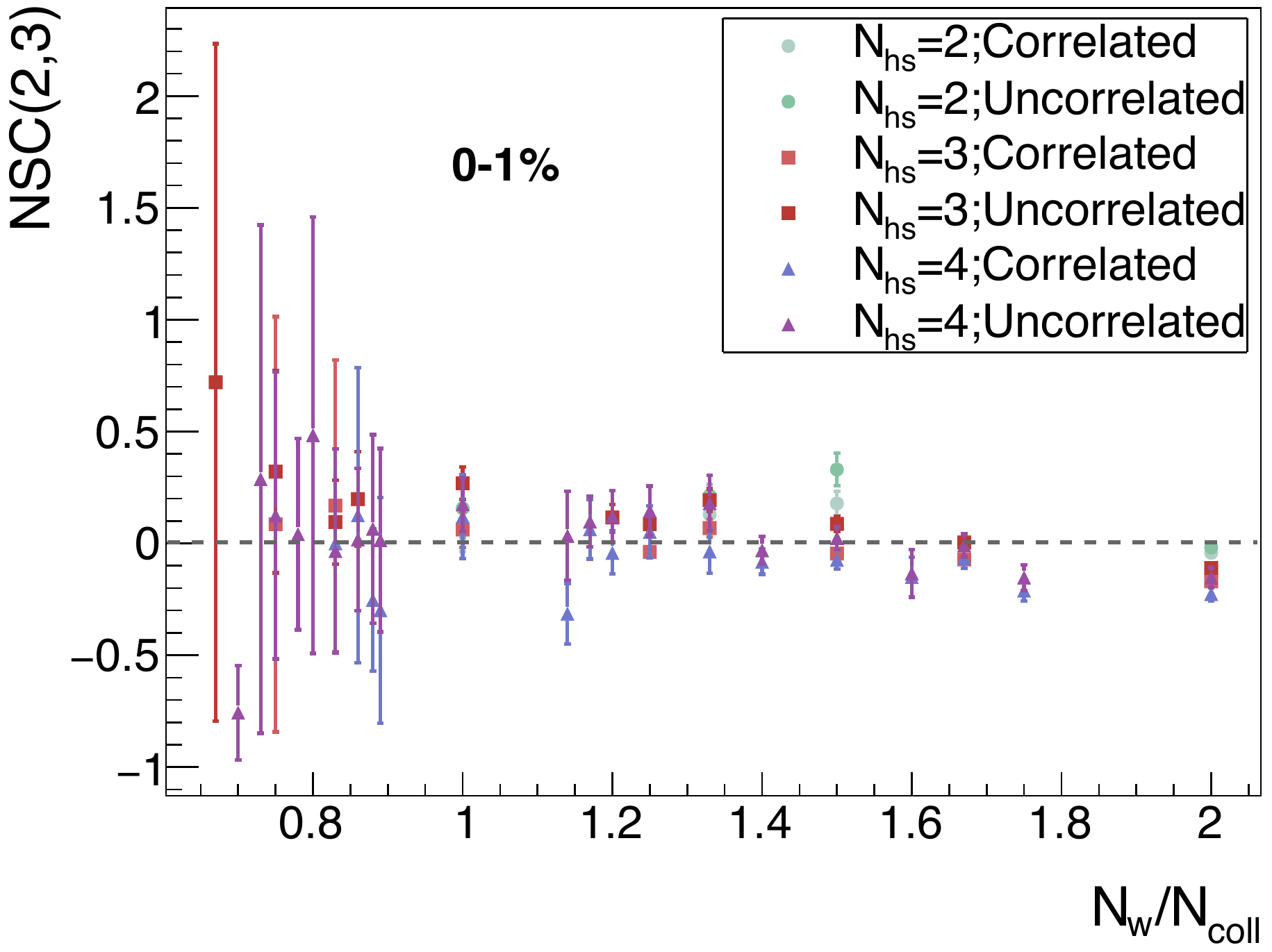}
\end{minipage}
\vspace*{-0.25cm}
\caption{Left: Average value of NSC(2,3) as a function of centrality for N$_{hs}\!=\!4$ in the correlated (red) and uncorrelated (purple) scenarios. Right: Average value of NSC(2,3) as a function of N$_{w}\!/\!$N$_{\rm{coll}}$ for $2\!\leq\!$N$_{hs}$$\!\leq\!4$ with and without correlations. }
\label{nsc23_param_plot}
\end{figure}

At this point it is natural to wonder whether the negative sign of NSC$(2,3)$ is a unique feature of the correlated scenario. To check this hypothesis it is necessary to explore the parameter space of our model as it is done in the left pannel of Fig.~\ref{nsc23_param_plot} where the number of hot spots is increased to 4. Although the qualitative effect of the spatial correlations persist, i.e. the correlated curve is always below the uncorrelated scenario in the highest centrality bins, an important comment is in order: NSC$(2,3)$ is compatible with negative values, within statistical uncertainty, in the [$0\!-\!0.1\%$] bin for the uncorrelated case. Therefore, the interplay of the different scales $\lbrace{R_{hs},r_c,N_{hs}\rbrace}$ is decisive in the sign of NSC$(2,3)$ within our framework.

Finally, we study the dependence of NSC$(2,3)$ on $N_{w}/N_{\rm coll}$ for the highest centrality bins with and without correlations and varying the number of hot spots. The results are shown on the right pannel of Fig.~\ref{nsc23_param_plot}. Remarkably, the value of the cumulant is the same in all the different scenarios for a given value of $N_{w}/N_{\rm coll}$. Nor the presence/absence of correlations neither the number of constituent hot spots modify the value of NSC$(2,3)$ for a fixed $N_{w}/N_{\rm coll}$. This feature backs up the idea that $N_{w}/N_{\rm coll}$ is a potential candidate to be the critical parameter controlling the sign of NSC$(2,3)$.

All in all, the aforementioned results plus the ones presented in \cite{Albacete:2017ajt} lead us to reach the firm conclusion that NSC$(2,3)$ is extremely sensitive to the initial state fluctuations and can help to discriminate between different parameterizations of the proton geometry.

\section{Outlook: hydrodynamic evolution}
The natural continuation of this work is to check the flow harmonic coefficients ($v_n$) are affected too by the inclusion of spatial correlations between the gluonic hot spots.  

To answer this question we use a 2+1D viscous hydrodynamic setup, thoroughly described in~\cite{Niemi:2015qia}, that can be summarized as follows. First, the simulation is initialized with the resulting entropy profiles from the MC-Glauber described in Section~\ref{model} at proper time $\tau\!=\!0.2$~fm. Further, both the shear-stress tensor ($\pi_{\mu\nu}$) and the transverse velocity are set to zero. The parametrization of the equation of state is obtained from~\cite{Huovinen:2009yb}, and the decoupling temperature is chosen to be $T_{\rm dec}\!=\!100$~MeV. Regarding the transport coefficients, both the heat conductivity and the bulk viscosity are neglected, while the temperature dependence of $\eta/s$ is modeled as in \cite{Niemi:2015qia} with a minimum at $T\!=\!150$~MeV.
 
\begin{figure}
\begin{minipage}{.5\textwidth}
  \centering
\includegraphics[width=1\textwidth]{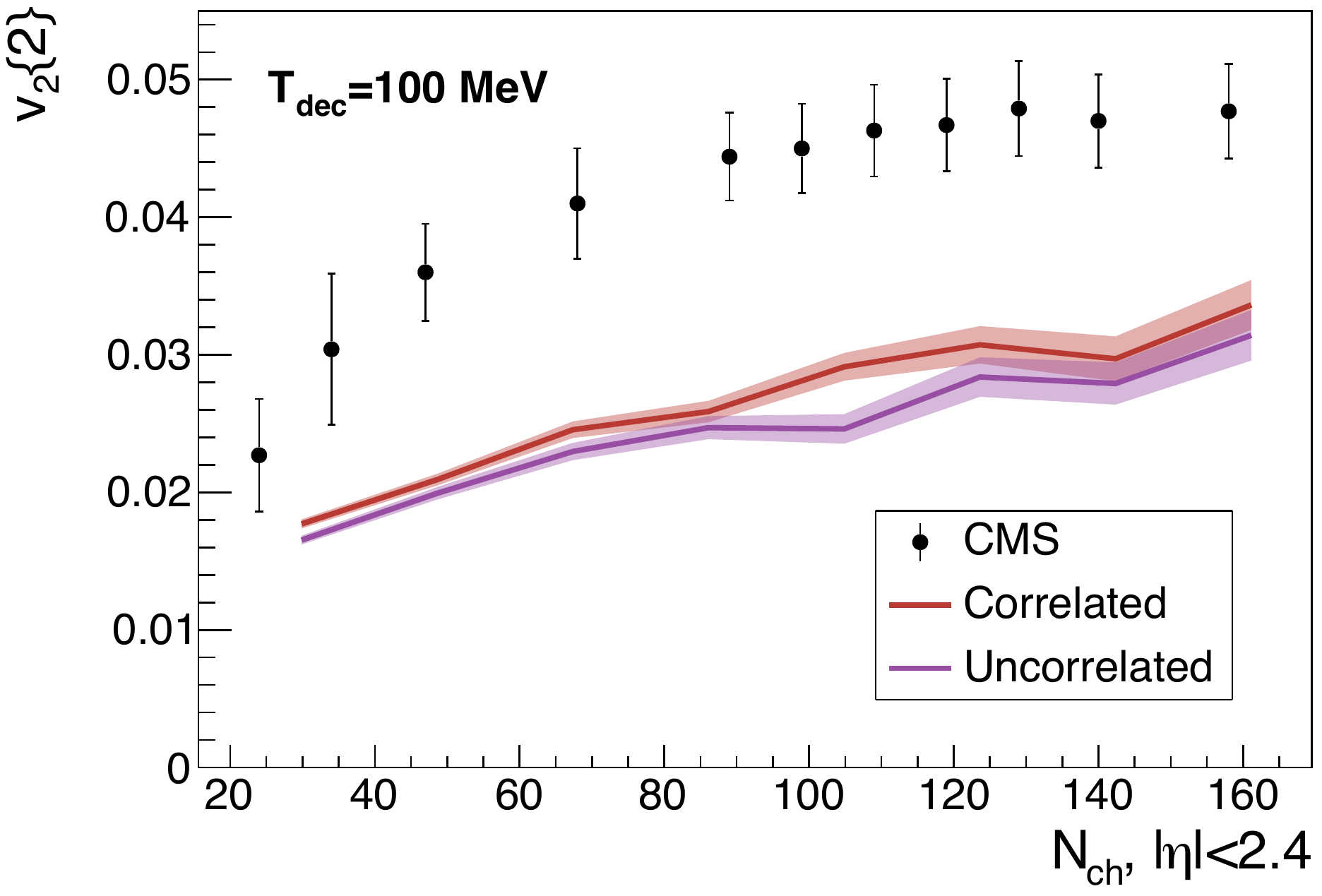}
\end{minipage}%
\hspace{0.2cm}
\begin{minipage}{.5\textwidth}
  \centering
\includegraphics[width=1\textwidth]{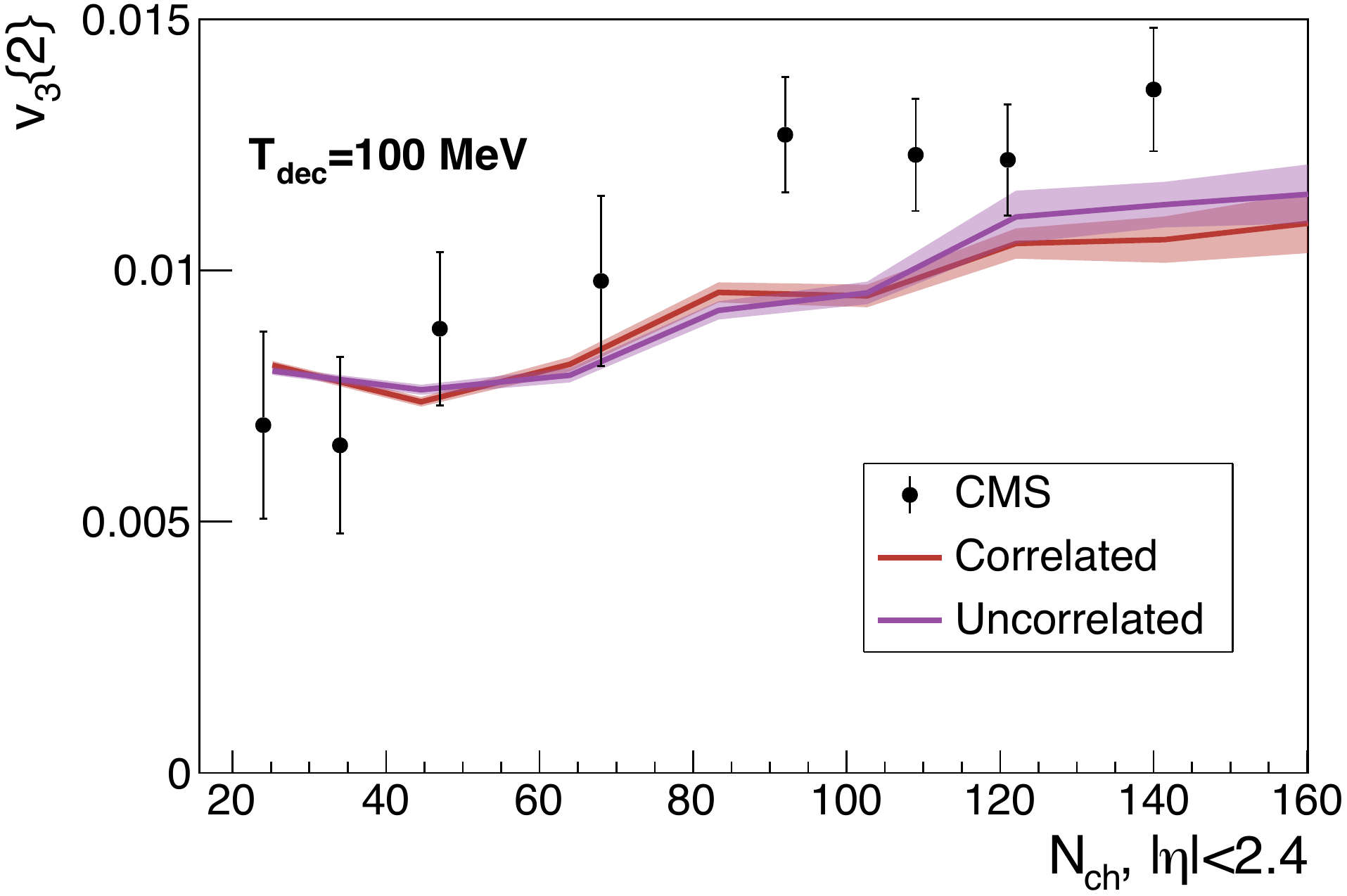}
\end{minipage}
\vspace*{-0.25cm}
\caption{Elliptic (left) and triangular (right) flow coefficients for the correlated (red line) and uncorrelated (purple line) cases as a function of the number of charged particles compared to the CMS data \cite{Khachatryan:2016txc}. The colored bands indicate the statistical uncertainty. }
\label{flowplot}
\end{figure}

Preliminary results on the elliptic ($v_2$) and triangular ($v_3$) flow for charged particles as a function of the multiplicity are depicted in Fig.~\ref{flowplot}~\cite{working}. Noticeably, the imprints of the spatial correlations are visible in the case of the elliptic flow that is enhanced in the correlated scheme as it was the case for the  eccentricity~\cite{Albacete:2016gxu}. In turn, $v_3$, with the current statistical precision, is compatible in both scenarios. Improving the quantitative description of the data by scanning the parameter space of our model and confirming the sensitivity of $v_2$ to a detailed description of the proton substructure are the two main lines of research that we will pursue in upcoming publications. 

\section*{Acknowledgements}This work was partially supported by a Helmholtz Young Investigator Group VH-NG-822 from the Helmholtz Association and GSI, a FP7-PEOPLE-2013-CIG Grant of the European Commission, reference QCDense/ 631558, by Ram\'on y Cajal and MINECO projects reference RYC-2011-09010 and FPA2013-47836 and by the DFG through the grant CRC-TR 211. HN is supported by the Academy of Finland, project 297058. We acknowledge the CSC–IT Center for Science in Espoo, Finland, for the allocation of the computational resources.



%% References


\bibliographystyle{elsarticle-num}
\bibliography{proc_QM18}






% \end{thebibliography}

\end{document}